\documentclass{article}
\usepackage{csquotes}
\usepackage{spconf,amsmath,graphicx,siunitx,microtype,amsfonts}
\usepackage[numbers]{natbib}
\usepackage{array}
\usepackage{booktabs}
\usepackage{microtype}
\usepackage{url}
\usepackage{amsmath}
\usepackage{siunitx}
\usepackage{color}
\usepackage{multirow}
\usepackage{booktabs}
\ninept

\title{Simple Pooling Front-ends For Efficient Audio Classification}

\name{
      Xubo Liu$^{1}$,
      Haohe Liu$^{1}$,
      Qiuqiang Kong$^{2}$,
      Xinhao Mei$^{1}$, 
      Mark D. Plumbley$^{1}$,
      Wenwu Wang$^{1}$
      }
      
\address{$^1$Centre for Vision, Speech and Signal Processing~(CVSSP), University of Surrey, UK\\
  $^2$Speech, Audio, and Music Intelligence~(SAMI) Group, ByteDance, China\\
 }

\begin{document}

\maketitle

\begin{abstract}
Recently, there has been increasing interest in building efficient audio neural networks for on-device scenarios. Most existing approaches are designed to reduce the size of audio neural networks using methods such as model pruning. In this work, we show that instead of reducing model size using complex methods, eliminating the temporal redundancy in the input audio features (e.g., mel-spectrogram) could be an effective approach for efficient audio classification. To do so, we proposed a family of simple pooling front-ends (SimPFs) which use simple non-parametric pooling operations to reduce the redundant information within the mel-spectrogram. We perform extensive experiments on four audio classification tasks to evaluate the performance of SimPFs. Experimental results show that SimPFs can achieve a reduction in more than half of the number of floating point operations (FLOPs) for off-the-shelf audio neural networks, with negligible degradation or even some improvements in audio classification performance.
\end{abstract}
\section{Introduction}
\label{sec:intro}
Audio classification is an important research topic in the field of signal processing and machine learning. There are many applications of audio classification, such as acoustic scene classification \cite{sun2022deep}, sound event detection \cite{mesaros2021sound} and keywords spotting \cite{warden2018speech}. Audio classification plays a key role in many real-world applications including acoustic monitoring \cite{radhakrishnan2005audio}, healthcare \cite{peng2009healthcare} and multimedia indexing \cite{kiranyaz2006generic}.

Neural network methods such as convolutional neural networks (CNNs) have been used for audio classification and have achieved state-of-the-art performance \cite{kong2020panns, gong2021psla, xu2018large}. Generally, state-of-the-art audio classification models are designed with large sizes and complicated modules, which make the audio classification networks computationally inefficient, in terms of e.g. the number of floating point operations (FLOPs) and running memory. However, in many real-world scenarios, audio classification models need to be deployed on resource-constrained platforms such as mobile devices \cite{xiao2022continual}. 

There has been increasing interest in building efficient audio neural networks in the literature. Existing methods can generally be divided into three categories. The first is to utilize model compression techniques such as pruning \cite{singh2022passive, singh2022low}. The second is to transfer the knowledge from a large-scale pre-trained model to a small model via knowledge distillation \cite{futami2020distilling, lu2017knowledge, choi2022temporal}. The last one is to directly exploit efficient networks for audio classification, such as MobileNets \cite{kong2020panns, liang2020channel}. In summary, these methods mainly focus on reducing model size. However, the computational cost (e.g., FLOPs) of the audio neural network is not only determined by the size of the model but also highly dependent on the size of the input features. 

\begin{figure}[htbp]
  \includegraphics[page=1,width=1.0\linewidth]{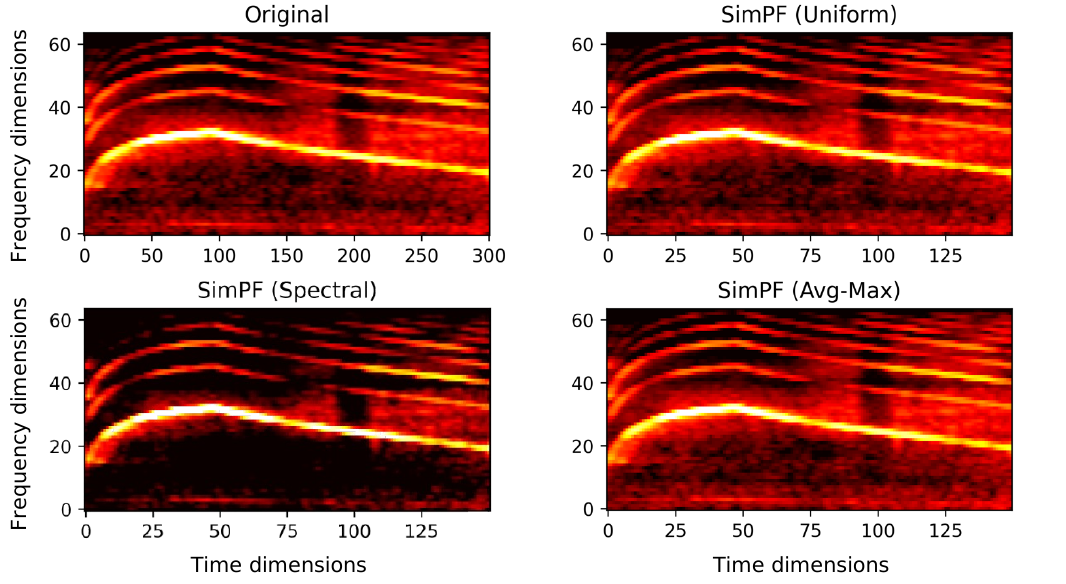}
    \caption{Visualization of the impact of SimPFs on the mel-spectrogram of a siren audio clip with ~50\% compression factor.}
    \label{fig-spec-compare}
\end{figure}

As existing audio neural networks usually take mel-spectrogram which may be temporally redundant. For example, the pattern of a siren audio clip is highly repetitive in the spectrogram, as shown in Figure \ref{fig-spec-compare}. In principle, if one can remove the redundancy in the input mel-spectrogram, the computational cost can be significantly reduced. However, reducing input feature size for audio neural networks has received little attention in the literature, especially in terms of improving their computation efficiency. 

In this paper, we propose a family of \textbf{sim}ple \textbf{p}ooling \textbf{f}ront-ends (SimPFs) for improving the computation efficiency of audio neural networks. SimPFs utilize simple non-parametric pooling methods (e.g., max pooling) to eliminate the temporally redundant information in the input mel-spectrogram. The simple pooling operation on an input mel-spectrogram achieves a substantial improvement in computation efficiency for audio neural networks. To evaluate the effectiveness of SimPFs, we conduct extensive experiments on four audio classification datasets including DCASE19 acoustic scene classification \cite{mesaros2018multi}, ESC-50 environmental sound classification \cite{piczak2015esc}, Google SpeechCommands keywords spotting \cite{warden2018speech}, and AudioSet audio tagging \cite{gemmeke2017audio}. We demonstrate that SimPFs can reduce more than half of the computation FLOPs for off-the-shelf audio neural networks \cite{kong2020panns}, with negligibly degraded or even improved classification performance. For example, on DCASE19 acoustic scene classification, SimPF can reduce the FLOPs by 75\% while improving the classification accuracy approximately by 1.2\%. Our proposed SimPFs are simple to implement and can be integrated into any audio neural network at a negligible computation cost. The code of our proposed method is made available at GitHub\footnote{\url{https://github.com/liuxubo717/SimPFs}}.

The remainder of this paper is organized as follows. The next section introduces the related work of this paper. Section \ref{sec:method} introduces the method SimPFs we proposed for efficient audio classification. Section \ref{sec:exp} presents the experimental settings and the evaluation results. Conclusions and future directions are given in Section \ref{sec:conclusion}.

\section{Related work}
\label{sec:related_works}
Our work relates to several works in the literature: efficient audio classification, feature reduction for audio classification, and audio front-ends. We will discuss each of these as follows.

\subsection{Efficient audio classification}
Efficient audio classification for on-device applications has attracted increasing attention in recent years. Singh et al. \cite{singh2022passive,singh2022low,singh2020svd} proposed to use pruning method to eliminate redundancy in audio convolutional neural networks for acoustic scene classification, which can reduce approximately 25\% FLOPs at 1\% reduction in accuracy. Knowledge distillation methods \cite{futami2020distilling, lu2017knowledge, choi2022temporal} have been used for efficient audio classification via transferring knowledge of large teacher models to small student on-device models. Efficient models such as MobileNets \cite{howard2017mobilenets} have been proposed for visual applications to mobile devices. Kong et al. \cite{kong2020panns} have adapted MobileNet for audio tagging, demonstrating its potential to improve computational efficiency for audio classification. Unlike these methods, which focus on reducing model size, our proposed SimPF aims to reduce the size of input features.  

\subsection{Feature reduction for audio classification}
Feature reduction methods such as principal component analysis (PCA) have been widely investigated for audio classification with classical machine learning methods such as discriminative support vector machines (SVMs) \cite{geiger2013large, zubair2013dictionary}. The most relevant work to SimPFs in the literature is \cite{zubair2013dictionary}, where max and average pooling operations are applied to sparse acoustic features to improve the performance of SVM-based audio classification, especially in a noisy environment. In contrast to this method, SimPFs are designed to improve the efficiency of audio neural networks, whose computational cost is highly dependent on the size of the input features. In addition, the effectiveness of SimPFs is extensively evaluated on various audio classification benchmarks.
\vspace{-1em}
\subsection{Audio Front-ends}
Audio front-ends were studied as an alternative to mel-filterbanks for audio classification in the last decade. For example, trainable front-ends SincNet \cite{ravanelli2018speaker} and LEAF \cite{zeghidour2021leaf} are proposed for learning audio features from the waveform. These front-ends perform better than using traditional mel-filterbanks on various audio classification tasks. Unlike existing work on learnable front-ends, SimPFs are non-parametric and built on top of a widely-used mel-spectrogram for audio neural networks. Our motivation for designing SimPFs is not to learn a replacement for mel-spectrogram, but to eliminate temporal redundancy in mel-spectrograms. This redundancy significantly impacts on the efficiency of audio neural networks but is often ignored by audio researchers. 
\begin{table*}[ht]
	\centering
		\begin{tabular}{@{}c|ccc|cc|cc@{}}
			\toprule
			Model (CNN10 \cite{kong2020panns})                       & \multicolumn{3}{c|}{DCASE19}                                                                                 & \multicolumn{2}{c|}{ESC-50}                                                                                            & \multicolumn{2}{c}{SpeechCommands}                         \\ \midrule
			Baseline                       & \multicolumn{3}{c|}{0.710}                                                                                               & \multicolumn{2}{c|}{0.850}                                                                                               & \multicolumn{2}{c}{0.971}                                     \\ \midrule
			\multirow{2}{*}{Front-end} & \multicolumn{3}{c|}{Compression Factor $k$}                                                                       & \multicolumn{2}{c|}{Compression Factor $k$}                                                                       & \multicolumn{2}{c}{Compression Factor $k$}             \\ \cmidrule(l){2-8} 
			                                            & \multicolumn{1}{c}{50\%}           & \multicolumn{1}{c}{25\%}           & 10\%             & \multicolumn{1}{c}{50\%}           & 25\%             & \multicolumn{1}{c}{50\%}           & \multicolumn{1}{c}{25\%} \\ \midrule
			$\text{SimPF (Max)}$        & \multicolumn{1}{c}{\textbf{0.722}} & \multicolumn{1}{c}{\textbf{0.722}} & 0.706           & \multicolumn{1}{c}{\textbf{0.853}} & 0.828 &  \multicolumn{1}{c}{0.969}          & 0.963                    \\ \midrule
			$\text{SimPF (Avg)}$               & \multicolumn{1}{c}{\textbf{0.721}} & \multicolumn{1}{c}{\textbf{0.721}} & 0.701 &  \multicolumn{1}{c}{0.846}          & 0.831  & \multicolumn{1}{c}{\textbf{0.971}} & 0.964                    \\ \midrule
			$\text{SimPF (Avg-Max)}$   & \multicolumn{1}{c}{\textbf{0.724}} & \multicolumn{1}{c}{\textbf{0.720}} & 0.705  & \multicolumn{1}{c}{0.849}          & 0.823  & \multicolumn{1}{c}{\textbf{0.971}} & 0.961                    \\ \midrule
			$\text{SimPF (Spectral)}$   & \multicolumn{1}{c}{\textbf{0.727}} & \multicolumn{1}{c}{\textbf{0.722}} & 0.709          & \multicolumn{1}{c}{0.845}          & 0.821 &  \multicolumn{1}{c}{0.970}          & 0.964                    \\ \midrule
			$\text{SimPF (Uniform)}$    & \multicolumn{1}{c}{\textbf{0.718}} & \multicolumn{1}{c}{\textbf{0.712}} & 0.682           & \multicolumn{1}{c}{0.846}          & 0.819  & \multicolumn{1}{c}{\textbf{0.971}} & 0.961                    \\ \bottomrule
		\end{tabular}
	\caption{CNN10 evaluation results on DCASE 2019 Task1, ESC-50, and SpeechCommands datasets. Baseline indicates the CNN10 model without SimPFs. The accuracy values where SimPFs outperform or perform the same as the CNN10 baseline are in bold.}
\end{table*}
\vspace{-1em}

\section{Proposed Method}
\label{sec:method}
 Mel-spectrogram is widely used as an input feature for neural network-based audio classification. Given an audio signal $x$, its mel-spectrogram is a two-dimensional time-frequency representation denoted as $X \in \mathbb{R}^{F \times T}$, where $T$ and $F$ represent the number of time frames and the dimension of the spectral feature, respectively. An audio neural network takes $X$ as the input and predicts the category $y$ of the input audio:
\begin{equation}
	\label{eq-1}
	\operatorname{g}(X; \theta) \mapsto y
\end{equation}
where $\operatorname{g}(\cdot, \theta)$ stands for the model parameterized by $\theta$. Generally, the computation cost of the neural network $\operatorname{g}$ is dependent on both the size of the parameter $\theta$ and the size of input $X$.

In this work, we propose to use simple non-parametric pooling methods to eliminate the temporal redundancy in the input mel-spectrogram. SimPFs can significantly improve the computational efficiency of audio neural networks without any bells and whistles. Formally, SimPFs take a mel-spectrogram $X$ as input and output a compressed time-frequency representation $C \in \mathbb{R}^{F \times kT}$, where $k \in (0, 1)$ is the compression coefficient in time domain and $\frac{1}{k}$ should be a positive integer. We will introduce a family of SimPFs which uses five pooling methods. 

\noindent\textbf{$\textit{SimPF (Max)}$} We apply a 2D max pooling with kernel size $(1, \frac{1}{k})$ over an mel-spectrogram $X$. The output is described as follows:
\begin{equation}
	\label{eq-1}
	C(f,t) = \mathop{\max\limits_{n=0,...,\frac{1}{k}-1}}X(f, \frac{t}{k} + n)
\end{equation}
where $f=0,...,F-1$ and $t=1,...,kT-1$.

\noindent\textbf{$\textit{SimPF (Avg)}$} Similar to SimPF (Max), we apply a 2D average pooling with kernel size $(1, \frac{1}{k})$ over an input mel-spectrogram $X$. Formally, the output is described as:
\begin{equation}
	\label{eq-1}
	C(f,t) = k\sum_{n=0}^{\frac{1}{k}-1}X(f, \frac{t}{k}  + n).
\end{equation}

\noindent\textbf{$\textit{SimPF (Avg-Max)}$} In this case, we add the outputs of $\textit{SimPF (Max)}$ and $\textit{SimPF (Avg)}$, which is defined as:

\begin{equation}
	\begin{aligned}
		\label{eq-1}
		C(f,t) = & \mathop{\max\limits_{n=0,...,\frac{1}{k}-1}}X(f, \frac{t}{k} + n) + k\sum_{n=0}^{\frac{1}{k}-1}X(f, \frac{t}{k}  + n).
	\end{aligned}
\end{equation}

\noindent\textbf{$\textit{SimPF (Spectral)}$} We adapt the spectral pooling method proposed in \cite{rippel2015spectral}. Concretely, the Discrete Fourier Transform (DFT) $y$ of the input mel-spectrogram $X$ is computed by:

\begin{equation}
	\begin{aligned}
		\label{eq-1-1}                               
		y = \operatorname{DFT}(X) \in \mathbb{C}^{F \times T}
	\end{aligned}
\end{equation}
 and the zero frequency is shifted to the center of $y$. Then, a bounding box of size $(F, kT)$ crops $y$ around its center to produce $y^{\rm{crop}} \in \mathbb{C}^{F \times kT}$. The output is obtained by exerting inverse DFT on $y^{\rm{crop}}$: 

\begin{equation}
	\begin{aligned}
		\label{eq-1}                               
		C = \operatorname{DFT}^{\operatorname{inverse}}(y^{\rm{crop}}).
	\end{aligned}
\end{equation}

\noindent\textbf{$\textit{SimPF (Uniform)}$} We uniformly sample one spectral frame every $\frac{1}{k}$ frames. The output of is calculated by:
\begin{equation}
	\label{eq-1}
	C(f,t) = X(f,\frac{t}{k}).
\end{equation}

We visualize the mel-spectrogram of a siren audio clip and the compressed spectrograms using different SimPFs with 50\% compression factor in Figure 1. Intuitively, we can observe that even though the resolution of the spectrogram is compressed by half, the pattern of the siren remains similar in the spectrogram, which indicates high redundancy in the siren spectrogram.

\vspace{-3em}
\section{Experiments and Results}
\label{sec:exp}

\subsection{Datasets}
\noindent \textbf{DCASE 2019 Task 1} \cite{tau2019} is an acoustic scene classification task, with a development set consisting of \num{10}-second audio clips from \num{10} acoustic scenes such as airport and metro station. 
In the development set, \num{9185} and \num{4185} audio clips are used for training and validation, respectively. We will refer to this dataset as \textit{DCASE19}.

\noindent \textbf{ESC-50} \cite{piczak2015esc} consists of \num{2000} five-second environmental audio clips. ESC-50 is a balanced dataset with \num{50} sound categories, including animal sounds, natural soundscapes, human sounds (non-speech), and ambient noises. Each sound class has \num{40} audio clips. The dataset is pre-divided into five folds for cross-validation.

\noindent 
\textbf{SpeechCommands} \cite{warden2018speech} contains \num{65}K speech utterances from various speakers. Each utterance is one second long and belongs to one of \num{30} classes corresponding to a speech command such as \enquote{Go}, \enquote{Stop}, \enquote{Left}, and
\enquote{Down}. We divided the datasets by a ratio of \num{80}:\num{10}:\num{10} for training, validation, and testing, respectively.

\noindent \textbf{AudioSet} \cite{gemmeke2017audio}
is a large-scale audio dataset with \num{527} sound classes in total. The audio clips are sourced from YouTube videos. The training set consists of \num{2063839} audio clips. The evaluation set has \num{85487} test clips. We convert all audio clips to monophonic and pad the audio clips to ten seconds with silence if they are shorter than ten seconds.

\subsection{Experiment setup}

\textbf{Baseline systems}~
We evaluate our proposed approach using several off-the-shelf audio classification methods proposed in \cite{kong2020panns}. As for the evaluation of ESC-50, DCASE19, and SpeechCommands dataset, we use two baseline models, CNN10 and MobileNetV2. On AudioSet, we conduct the experiment on CNN14 and MobileNetV2. CNN10 and CNN14 are both large-scale audio neural networks, and MobileNetV2 is designed with low complexity by multiply-add operations and fewer parameters. Hence, MobileNetV2 is suitable for on-device scenarios. We train all the models from scratch.

\noindent
\textbf{Implementation details}~ We load the audio clips using the sampling rate as provided in the original dataset. The audio clip is converted to \num{64}-dimensional log mel-spectrogram by the short-time Fourier transform with a window size of \num{1024} samples, a hop size of \num{320} samples, and a Hanning window. The baseline audio classification networks are optimized with the Adam optimizer with the learning rate \num{1e-3}. The batch size is set to \num{32} and the number of epochs is \num{300}, except for AudioSet where we run \num{15} epochs. Following ~\citep{gong2021psla}, random SpecAugment \cite{park2020specaugment} is used for data augmentation.

\noindent
\textbf{Evaluation metrics}~
Following ~\citep{zeghidour2021leaf}, we use accuracy as the evaluation metric on ESC-50, DCASE19, and SpeechCommands datasets. As for the AudioSet dataset, we use mean average precision (mAP) to evaluate the performance of audio tagging.
\vspace{-1em}
\begin{table}[t]
\centering
\begin{tabular}[\linewidth]{c c c c c} 
 \hline
Models & Baseline & SimPF (50\%) & SimPF (25\%) \\ 
 \hline
 MobileNetV2 & 488.78M & 243.80M & 121.33M \\ 
  CNN10 & 19.55G & 9.76G & 4.85G \\ 
 CNN14 & 30.04G & 14.97G & 7.41G \\ 
 \hline
\end{tabular}
	\caption{FLOPs analysis for CNN10, MobileNetV2, and CNN14 baseline models. The FLOPs are computed for one 10-second audio clip with a 44 kHz sampling rate. }
\label{table-2}
\vspace{-1.5em}
\end{table}

\begin{table*}[ht]
	\centering
		\begin{tabular}{@{}c|ccc|cc|cc@{}}
			\toprule
			Model (MobileNetV2 \cite{kong2020panns})                       & \multicolumn{3}{c|}{DCASE19}                                                                                 & \multicolumn{2}{c|}{ESC-50}                                                                                            & \multicolumn{2}{c}{SpeechCommands}                         \\ \midrule
			Baseline                       & \multicolumn{3}{c|}{0.670}                                                                                               & \multicolumn{2}{c|}{0.779}                                                                                               & \multicolumn{2}{c}{0.969}                                     \\ \midrule
			\multirow{2}{*}{Front-end} & \multicolumn{3}{c|}{Compression Factor $k$}                                                                       & \multicolumn{2}{c|}{Compression Factor $k$}                                                                       & \multicolumn{2}{c}{Compression Factor $k$}             \\ \cmidrule(l){2-8} 
			                                            & \multicolumn{1}{c}{50\%}           & \multicolumn{1}{c}{25\%}           & 10\%             & \multicolumn{1}{c}{50\%}           & 25\%             & \multicolumn{1}{c}{50\%}           & \multicolumn{1}{c}{25\%} \\ \midrule
			$\text{SimPF (Avg-Max)}$       & \multicolumn{1}{c}{0.660} & \multicolumn{1}{c}{\textbf{0.682}} & 0.661           & \multicolumn{1}{c}{\textbf{0.784}} & 0.764          & 0.956 & -                   \\ \midrule
			$\text{SimPF (Spectral)}$                 & \multicolumn{1}{c}{\textbf{0.673}} & \multicolumn{1}{c}{0.668} & 0.667 &  \multicolumn{1}{c}{\textbf{0.785}}          & 0.772   & 0.953 & -                    \\ 
 \bottomrule
		\end{tabular}
	\caption{MobileNetV2 evaluation results on DCASE 2019 Task1, ESC-50, and SpeechCommands datasets. The accuracy values where SimPFs outperform the MobileNetV2 baseline are in bold.}
	  \vspace{-1.5em}

\end{table*}
\begin{table}[]
	\centering
		\begin{tabular}{@{}c|cc@{}}
			\toprule
			Model (CNN14 \cite{kong2020panns})                       & \multicolumn{2}{c}{AudioSet}                                                                                          \\ \midrule
			Baseline                       & \multicolumn{2}{c}{0.432}                                                                                                                   \\ \midrule
			\multirow{2}{*}{Front-end} & \multicolumn{2}{c}{Compression Factor $k$}                                                                               \\ \cmidrule(l){2-3} 
			                                            & \multicolumn{1}{c}{50\%}           & \multicolumn{1}{c}{25\%}                     \\ \midrule
			$\text{SimPF (Avg-Max)}$       & \multicolumn{1}{c}{0.424} & \multicolumn{1}{c}{0.400}                              \\ \midrule
			$\text{SimPF (Spectral)}$                & \multicolumn{1}{c}{0.426} & \multicolumn{1}{c}{0.397}                      \\ 
 \bottomrule
		\end{tabular}
	\caption{CNN14 evaluation results on AudioSet.}
\end{table}
\begin{table}[h]
	\centering
		\begin{tabular}{@{}c|cc@{}}
			\toprule
			Model (MobileNetV2 \cite{kong2020panns})                       & \multicolumn{2}{c}{AudioSet}                                                                                          \\ \midrule
			Baseline                       & \multicolumn{2}{c}{0.318}                                                                                                                   \\ \midrule
			\multirow{2}{*}{Front-end} & \multicolumn{2}{c}{Compression Factor $k$}                                                                               \\ \cmidrule(l){2-3} 
			                                            & \multicolumn{1}{c}{50\%}           & \multicolumn{1}{c}{25\%}                     \\ \midrule
			$\text{SimPF (Avg-Max)}$       & \multicolumn{1}{c}{0.312} & \multicolumn{1}{c}{0.297}                              \\ \midrule
			$\text{SimPF (Spectral)}$                & \multicolumn{1}{c}{0.314} & \multicolumn{1}{c}{0.293}                      \\ 
 \bottomrule
		\end{tabular}
	\caption{MobileNetV2 evaluation results on AudioSet.}
	  \vspace{-1.5em}

\end{table}

\subsection{Evaluation results and analysis}
\subsubsection{Computation cost analysis (FLOPs)}
We analyze the impact on FLOPs reduction of our SimPFs on compression coefficients 50\% and 25\% for three baseline systems. Table 2 shows the FLOPs of the model to infer a 10-seconds audio clip with a sampling rate of 44 kHz. For our three baseline models, the compression ratio on the input spectrogram is roughly equivalent to the FLOPs reduction ratio. We refer to the spectrogram compression ratio as the FLOPs reduction ratio in the later experiment analysis.
\vspace{-1em}

\subsubsection{DCASE19}
For the CNN10 model, we evaluate the effectiveness of all our proposed SimPF with three compression factors: 50\%, 25\%, and 10\% (as introduced in Section 3). The experimental results are shown in the left column in Table 1. Overall, \textit{SimPF (Spectral)} performs best among all SimPFs candidates on three compression coefficient settings. Even though MobileNetV2 is smaller than CNN10, we find that \textit{SimPF (Spectral)} can reduce the FLOPs by roughly 50\% and 25\% while still improving the classification accuracy by 1.7\% and 1.2\%, respectively. Even reducing the FLOPs by 90\%, the classification accuracy only drops by 0.01\%. For MobileNetV2, we evaluate the performance of two representative candidates \textit{SimPF (Avg-Max)} and \textit{SimPF (Spectral)} with three compression factor 50\%, 25\%, and 10\%. We find similar trends to CNN10 model, as shown in the left column of Table 3.  \textit{SimPF (Avg-Max)} improves the accuracy by 1.2\% on 25\% setting and \textit{SimPF (Spectral)} only sacrifices 0.3\% accuracy on 10\% setting. Experimental results on CNN10 and MobileNetV2 demonstrate the superior performance of our proposed SimPF for acoustic scene classification, also indicating 
the highly redundant information in the acoustic scene data.
\vspace{-1em}
\subsubsection{ESC-50}
For CNN10 model, we evaluate the performance of all our proposed SimPF methods with two compression factors: 50\% and 25\%. The experimental results are shown in the middle column of Table 1. On the 50\% setting, the best candidate \textit{SimPF (Max)} achieves the accuracy improvement by 0.3\%. On the 25\% setting, the best candidate \textit{SimPF (Avg)} reduces an accuracy by 1.9\%. For MobileNetV2 model, we evaluate the performance of two representative candidates \textit{SimPF (Avg-Max)} and \textit{SimPF (Spectral)} with two compression factor 50\%, 25\%, as shown in the middle column of Table 2. \textit{SimPF (Spectral)} performs better in these two different compression coefficient settings. Specifically, \textit{SimPF (Spectral)} improves the classification accuracy by 0.6\% on the 50\% setting, and slightly decreases the accuracy by 0.7\% on the 25\% setting. The performance gain of SimPFs on ESC-50 is not as good as that on DCASE19 but is still decent in terms of the trade-off between the accuracy and FLOPs.

\vspace{-1em}
\subsubsection{SpeechCommands}
For CNN10 model, we evaluate all our proposed SimPFs with two compression factors 50\% and 25\%. The results are shown in the right column of Table 1. On the 50\% setting, \textit{SimPF (Avg)}, \textit{SimPF (Avg-Max)}, \textit{SimPF (Max)}, and \textit{SimPF (Uniform)} achieve the equivalent performance as the baseline system. On the 25\% setting, the best two candidates \textit{SimPF (Avg)} and \textit{SimPF (Spectral)} reduce the accuracy only by 0.7\%. For the MobileNetV2 model, we evaluate the performance of two representative candidates \textit{SimPF (Avg-Max)} and \textit{SimPF (Spectral)} with compression factor at 50\% (25\% setting is not available\footnote{The 25\% SimPF setting is not available as 25\% of one-second speech clip is too short for MobileNetV2 to process.}), as shown in the right column of Table 2. The best candidate \textit{SimPF (Avg-Max)} decreases the accuracy by 1.3\%. Evaluation results on SpeechCommands show that our proposed method is useful for short-utterance speech data. 

\vspace{-1em}

\subsubsection{AudioSet}
We evaluate the performance of two representative candidates \textit{SimPF (Avg-Max)} and \textit{SimPF (Spectral)} with two compression factors 50\% and 25\%, for the CNN10 model, as shown in Table 3. On the 50\% setting, \textit{SimPF (Spectral)} only reduces the mAP by 0.8\%, on the 25\% setting, \textit{SimPF (Avg-Max)} reduces the mAP by 3.2\%. Similar results we obtained for the MobileNetV2 model, as shown in Table 4. On the 50\% setting, \textit{SimPF (Spectral)} only reduces the mAP by 0.4\%, on the 25\% setting, \textit{SimPF (Avg-Max)} reduces the mAP by 2.1\%. SimPF can roughly reduce 50\% computation cost with a negligible mAP drop within 1\%. Even though tagging AudioSet is a more challenging task as compared with classification for other datasets, SimPFs achieve a promising trade-off between computation cost and mAP.

\vspace{-0.5em}
\section{Conclusion}
In this paper, we have presented a family of simple pooling front-ends (SimPFs) for efficient audio classification. SimPFs utilize non-parametric pooling methods (e.g., max pooling) to eliminate the temporally redundant information in the input mel-spectrogram. SimPFs achieve a substantial improvement in computation efficiency for off-the-shelf audio neural networks with negligible degradation or considerable improvement in classification performance on four audio datasets. In future work, we will study parametric pooling audio front-ends to adaptively reduce audio spectrogram redundancy.
\label{sec:conclusion}

\vspace{-0.5em}
\section{ACKNOWLEDGMENT}
\label{sec:ack}
This work is partly supported by UK Engineering and Physical Sciences Research Council (EPSRC) Grant EP/T019751/1 ``AI for Sound'', a Newton Institutional Links Award from the British Council (Grant number 623805725), British Broadcasting Corporation Research and Development~(BBC R\&D), a PhD scholarship from the Centre for Vision, Speech and Signal Processing (CVSSP), Faculty of Engineering and Physical Science (FEPS), University of Surrey, and a Research Scholarship from the China Scholarship Council (CSC). For the purpose of open access, the authors have applied a Creative Commons Attribution (CC BY) licence to any Author Accepted Manuscript version arising.
\bibliographystyle{IEEEtran}
\bibliography{strings,refs}

\end{document}